# Responsible Data Stewardship: Generative AI and the Digital Waste Problem

Vanessa Utz

Simon Fraser University

vutz@sfu.ca

## Abstract

As generative AI systems become widely adopted, they enable unprecedented creation levels of synthetic data across text, images, audio, and video modalities. While research has addressed the energy consumption of model training and inference, a critical sustainability challenge remains understudied: digital waste. This term refers to stored data that consumes resources without serving a specific (and/or immediate) purpose. This paper presents this terminology in the AI context and introduces digital waste as an ethical imperative within (generative) AI development, positioning environmental sustainability as core for responsible innovation. Drawing from established digital resource management approaches, we examine how other disciplines manage digital waste and identify transferable approaches for the AI community. We propose specific recommendations encompassing research directions, technical interventions, and cultural shifts to mitigate the environmental consequences of indefinite data storage. By expanding AI ethics beyond immediate concerns like bias and privacy to include intergenerational environmental justice, this work contributes to a more comprehensive ethical framework that considers the complete lifecycle impact of generative AI systems.

## Introduction

The explosive growth of generative AI has transformed how we create and consume digital content, enabling the rapid production of synthetic data across text, images, audio, and video modalities at record scale and speed. While this technological advancement has the potential to significantly increase human productivity, it also presents significant sustainability challenges that require urgent consideration.

Current discourse on AI sustainability has primarily centered on two resource-intensive processes: (1) the computational demands during the training and fine-tuning large models (Strubell, Ganesh & McCallum 2019), and (2) the energy required during inference as these systems generate outputs (Utz & DiPaola 2023). Researchers have quantified these impacts through carbon emissions (Lacoste et. al. 2019; Luccioni & Hernandez-Garcia 2023) and proposed various strategies to reduce them (Chien et. al. 2023). However, a critical dimension of generative AI's environmental footprint remains largely unexplored: the long-term consequences of storing generated data in perpetuity.

This gap in research warrants immediate investigation as generative AI adoption accelerates globally, with an estimated 500 million daily users as of 2024 (Qiang, Liu & Wang 2024). In this paper, we introduce the "digital waste" terminology to the AI community, including developers, academic researchers and end users, and position environmental sustainability as a core ethical consideration in responsible AI development.

Traditional AI ethics frameworks have predominantly focused on issues such as bias, fairness, privacy, and transparency (Prem 2023). However, as the climate crisis we are currently facing intensifies, the environmental footprint of AI-generated data deserves equal attention as an ethical concern of global significance. By framing digital waste as an ethical challenge with intergenerational implications, we establish sustainability as a core requirement rather than an optional consideration for AI-generated content.

Drawing parallels with digital waste management approaches from other fields, we introduce specific recommendations on how to tackle digital waste and aim to catalyze new thinking on how the AI community addresses these challenges. This work contributes to the growing field of sustainable AI research by broadening the scope beyond computational efficiency to encompass the entire lifecycle of AI-generated content and its potential impact on future generations.

## Understanding Digital Waste and its Material Reality

Digital waste, also referred to as data waste (Bietti & Vatanparast 2020), refers to stored data that does not serve a function, but still negatively affects the environment (i.e. due to emissions or resource extraction). Commonly encountered examples of digital waste include duplicates of files (such as vacation photos that are stored on several different devices, or within several locations on a single device) or old/outdated files that are no longer needed (such as old e-mails). This phenomenon occurs across the digital ecosystem, from individual consumer devices to massive corporate data centers, creating environmental impacts at multiple scales.

The concept extends beyond old emails and vacation photographs. It includes any redundant copies of the same data, outdated versions that remain stored alongside current iterations, temporary files that never get deleted, and data that has outlived its usefulness yet continues to occupy storage space. What makes digital waste particularly insidious is its seeming immateriality: unlike physical waste that accumulates visibly, digital waste accumulates silently, its environmental footprint largely unseen by end users. While this waste accumulation might not be immediately visible, digital waste still has physical substrates which have significant environmental impacts.

### Initial Hardware and Infrastructure Requirements

The physical infrastructure enabling data storage constitutes a complex global network of semiconductor chips, memory units, storage devices, and data centers. The environmental footprint of this infrastructure begins with manufacturing processes that require extensive resource extraction and refinement.

Semiconductor fabrication facilities produce memory chips, solid-state drives, and processing units that form the backbone of data storage systems. Production relies heavily on precious and rare earth metals including cobalt, gold, copper, and aluminum (Hosseini, Gao, Vivas-Valencia 2025). The extraction of these materials often involves destructive mining practices that lead to habitat destruction, soil erosion, and groundwater contamination.

Furthermore, water usage in semiconductor manufacturing has now reached alarming levels. Clean rooms require ultra-pure water for cleaning silicon wafers and cooling manufacturing equipment. This consumption has more than doubled over the past decade, with recent industry assessments indicating that chip manufacturing now consumes water nearly equivalent to the daily usage of 30 million Americans (Ruberti 2024). This consumption creates significant pressure on local water resources, particularly in already water-stressed communities.

The pollution footprint however extends beyond production facilities. Studies have documented heavy metal contamination, including tungsten, copper, and arsenic, in waterways downstream from semiconductor manufacturing facilities (Hsu et. al. 2011). These contaminants can persist in ecosystems for decades. The production of toxic waste associated with chip manufacturing has quadrupled over the last decade, reaching 874 kilotons in 2021 (Ruberti 2024).

The energy demands of hardware manufacturing are similarly substantial. A life-cycle analysis of the six leading global chip manufacturers revealed their combined annual energy consumption reached 27,768 GWh. This energy usage translated to approximately 16,369 kilotons of $CO_2$ equivalent emissions (a unit of measure for the global warming potential of all greenhouse gasses (GHGs)) (Ruberti 2023).

### Ongoing Data Center Maintenance

The environmental impact extends far beyond the initial manufacturing phase. Once operational, data centers create ongoing environmental burdens through their continuous operation. These facilities are designed for maximum reliability, which necessitates 24/7 operation with multiple redundancy systems (Monserrate 2022).

Modern data centers now consume more energy than some nations, with approximately 300 TWh required in 2021 (Alkrush et. al. 2024). Cooling systems represent the largest component of this energy use. Traditional air conditioning systems typically account for over 40% of a data center's total electricity consumption, driven by the significant heat generated by servers operating at capacity (Monserrate 2022).

To reduce energy consumption, many data center operators have begun implementing water-based cooling systems. While this approach significantly reduces electricity usage, it transfers the environmental burden from energy consumption to water usage. A typical small, water-cooled data center uses around 25 million liters of water annually (Mytton, 2021).

Additionally, pollution also poses a significant concern. Pollution associated with data centers manifests in two forms. 1) Noise pollution from cooling systems and backup generators affects workers and nearby communities and 2) the regular replacement cycle for servers and storage equipment generates substantial electronic waste. Enterprise-grade storage systems (i.e. GPUs in data centers) typically operate on 5-year replacement cycles (Horizon Technology 2024), creating a continuous stream of retired equipment containing hazardous materials.

The industry has made progress in some aspects of data center sustainability. Operational efficiency improvements and renewable energy adoption have helped reduce the energy intensity of data storage. However, these advances primarily address operating emissions rather than embodied carbon from manufacturing. As highlighted by Gupta et. al. (2021), hardware manufacturing and infrastructure construction continue to lag significantly in sustainability advances, compared to the operation phase.

## Generative AI and its Impact of Digital Waste

Generative AI represents a paradigm shift in how digital content is created and consumed, enabling the production of synthetic data at unprecedented rates. The relationship between generative AI and digital waste manifests through two interconnected mechanisms: the substantial resources con-sumed during content generation itself and the ongoing environmental burden of storing the resulting synthetic content indefinitely.

**Resource-Intensive Generation**
Creating synthetic data through generative AI systems is fundamentally resource-intensive, with computational demands varying based on the generation task and output modality. These systems rely on complex neural network architectures that perform numerous computationally expensive operations to transform input prompts into coherent outputs. Recent research by Luccioni et. al. (2024) has provided insights into the energy requirements of different generative and classification tasks. Generation tasks (such as text-to-image) consistently proved more energy and carbon-intensive than classification tasks due to the inherently greater complexity of media creation.

The modality of the generated content significantly influences resource requirements. Image-based tasks demand substantially more computational resources than text-only processes. Text-generation operations require a median of 0.042 kWh per 1000 inferences, which is equivalent to charging two smartphones (charging a modern average smartphone takes approximately 0.022 kWh). By contrast, image-generation tasks consume a median of 1.35 kWh per 1000 inferences, equivalent to charging approximately 61 smartphones (Luccioni et. al. 2024).

These energy demands intensify with newer multimodal systems. Novel video generation systems like OpenAI's Sora represent the frontier of resource intensity and are therefore understudied in their impact on energy consumption and emissions, initial work has shown that video-based generative AI systems require even more computational resources due to their iterative diffusion denoising process that is involved in creating moving imagery (Li, Jiang & Tiwari 2024).

Unlike traditional content creation, where human cognitive effort serves as a natural limiting factor, generative AI enables content creation at machine speed, allowing users to generate ever increasing volumes of data in decreasing time frames. Three factors specific to the user interfaces of generative AI systems further encourage large quantities of data generations: First, the iterative nature of interacting with generative systems leads users to create multiple versions while repeatedly refining prompts. A user might generate dozens of images before finding one that meets their requirements. Second, the ease of generation encourages exploratory creation without clear purpose. Third, synthetic content often lacks clear contextual metadata that might help users later determine its relevance, making it less likely to be deliberately assessed during storage cleanup.

This process also increases the volume of potentially disposable content, establishing the foundation for unprecedented levels of digital waste.

**Perpetual Storage Demands and Generational Debt**
The environmental footprint of generative AI extends far beyond the moment of content creation, due to the accumulation of digital waste files. Once generated, synthetic content requires continued storage either locally (i.e. on home computers) or on cloud services, which puts a strain on the physical resources required to manufacture and maintain the needed hardware and infrastructure. This perpetual storage requirement creates a "long tail" of environmental impact: an extended period during which the content continues to consume resources regardless of whether it serves an active purpose.

This persistent storage represents a form of intergenerational burden. When we store data indefinitely, we commit future generations to maintaining the infrastructure needed to support that storage, including ongoing energy consumption, hardware replacement, and resource extraction.

The scale of this potential burden becomes particularly concerning when we consider the current growth trajectory of generative AI adoption. Even if individual users only generate modest amounts of content that persists in storage, the cumulative environmental burden still grows significantly (Utz & DiPaola 2023). Unlike physical waste, which eventually degrades, digital waste can theoretically persist indefinitely unless deliberately deleted.

This persistence creates a sustainability debt that compounds overtime. Each generation that fails to address the accumulation of digital waste passes on a larger problem to subsequent generations, who must then devote increasingly significant resources to maintaining legacy data. This dynamic parallels other environmental challenges such as rising global temperatures, where delays in addressing root causes make ultimate solutions more difficult and costly. Despite these implications, this aspect of generative AI's environmental impact remains critically understudied.

# Digital Waste Management Approaches: Learning from Adjacent Fields

While digital waste presents a significant challenge for generative AI sustainability, it is not an entirely novel phenomenon. Other disciplines have developed frameworks and methodologies that can inform our approach to address digital waste in the context of AI. By examining how adjacent fields conceptualize and manage digital waste, we can identify and extract translational principles.

**Established Approaches to Digital Resource Management**
Several fields have developed systematic approaches to managing digital resources efficiently. For example, Digital Lean Manufacturing (DLM), which evolved from traditional manufacturing practices, offers insights into addressing digital waste.

Lean Manufacturing (LM) emerged in the 1990s as a formalized set of principles derived from the Toyota Production System (Bhamu & Sangwan 2014; Ciarniene & Vienazindiene 2012). At its core, it represents a systematic approach to identifying and eliminating waste, defined as any resource expenditure that does not create value for the end customer. As manufacturing operations have become increasingly digitalized through Industry 4.0 initiatives, Lean principles have evolved to also address digital resources (Powell & Romero 2021).

In DLM literature, practitioners distinguish between passive digital waste (missed opportunities to utilize existing data effectively) and active digital waste (issues arising from data overabundance) (Romero et. al. 2018). Passive waste includes scenarios where valuable data exists but remains unanalyzed or inaccessible. Active waste encompasses redundant data storage, excessive collection beyond what's needed, and maintaining outdated information. It is this active form of digital waste that we identify as a sustainability challenge for AI.

While digital waste has become a topic of discussion within DLM, for example the issue around storing and maintaining large quantities of generated data (Rossi et al 2022), only a limited number of works have attempted to actually tackle the inefficiencies and waste that are associated with these quantities of data. The majority of the literature in this area focuses on how data can be utilized to make processes more efficient. In the DLM and Industry 4.0 literature that does aim to address digital waste, the main challenge that has been identified is the problem of attempting to transfer waste reduction principles used for physical systems (i.e. the manufacturing of a physical product) for digital environments (Yarbrough, Harris & Purdy 2022). One strategy, however, that appears to translate from the physical to the digital paradigm, is Lean Thinking 4.0 (Rossi et. al. 2022). Lean Thinking 4.0 emphasizes establishing waste-conscious values and digital literacy before implementing any new technologies (Ciarniene & Vienazindiene 2012).

Information Lifecycle Management (ILM) offers another approach to addressing digital waste management. ILM focuses on managing information through defined stages from creation through archiving or deletion based on its changing value over time (Short 2007; Al-Fedaghi 2008). This approach emphasizes appropriate retention periods and automated processes to transition data through storage tiers as utility diminishes. These data governance frameworks from enterprise computing provide insights on establishing organizational responsibilities for data management. These frameworks typically include data quality assessment, metadata management, access controls, and systematic disposal protocols.

**Translational Principles for Generative AI**
Five key principles emerge from DLM and ILM approaches that transfer effectively to generative AI contexts:

1. Value-based assessment evaluates data based on its current and potential future utility rather than accumulating it indiscriminately. Central to both Lean Manufacturing and Information Management, this principle could guide frameworks for distinguishing between AI-generated content worth preserving and content that can be discarded, assessing not just immediate utility but also long-term value relative to storage costs.

2. Tiered storage architectures move data through different storage systems with varying performance characteristics based on access patterns and utility. Drawing from ILM practices, this could inform systems that automatically compress, or archive infrequently accessed outputs, or store generation parameters rather than full outputs for content that could be regenerated when needed.

3. Systematic pruning protocols to identify and remove data that no longer serves needs. Derived from Lean Manufacturing's emphasis on eliminating waste and ILM's structured approach to data retirement, this could guide the development of tools that help users identify and remove unnecessary AI-generated content, including automated identification of duplicates and content aging analysis.

4. Resource-conscious design creates systems with efficiency as a core consideration rather than an afterthought. This principle could influence generative AI interfaces that encourage thoughtful content creation rather than unlimited generation, such as interfaces that visualize environmental impact, default settings that limit unnecessary variations, and design patterns emphasizing quality over quantity.

5. Education before implementation establishes cultural foundations for responsible resource management before deploying new technologies. This principle, emphasized in Lean Thinking 4.0, recognizes that technical solutions alone cannot create sustainable practices without corresponding cultural norms, and could inform how organizations introduce generative AI tools.

These five principles provide a first foundation for developing sustainable approaches to generative AI by adapting established framework to digital waste management. The success of Lean Manufacturing in transforming resource management across diverse industries suggests that its core principles, properly adapted, could contribute significantly to addressing digital waste in AI contexts.

## Future Directions for Sustainable Data Practices

Having identified translatable principles from ILM and DLM, we now turn to specific recommendations targeting individual stakeholders for mitigating digital waste in generative AI. The environmental context within which we must address this sustainability challenge has now acquired an even higher level of urgency. With 2024 being recently confirmed as the hottest year on record since the beginning of temperature record keeping in the 1880s (Bardan 2025;

NOAA 2024), we have now reached 1.5-degree Celsius above the mid-19$^{th}$ century average, a threshold established in the Paris Agreement. The main driving factor between this increase in global temperatures are emissions of GHGs such as $CO_2$, CO and NO (Ramanathan & Feng 2009), and the effects of climate change are far reaching: A 2022 report on climate change impacts and vulnerabilities (Parmesan, Morecroft & Trisurat 2022) highlights concerning threats such as longer drought and wildfire seasons, fresh water supply loss, crop loss, extinctions within both fauna and flora, spreading of wildlife diseases, loss of settlements and infrastructure due to extreme weather events among others. As generative AI adoption accelerates globally, there is a pressing need to develop comprehensive strategies that address the sustainability implications of massive synthetic data generation and storage as ethical imperatives with intergenerational consequences. The responsible development and deployment of AI systems must therefore include consideration of their complete environmental footprint. The extensive timeframe of data storage requires particular attention. Current practices of indefinite data storage effectively commit future generations to maintaining digital infrastructure they had no role in creating and from which they may or may not derive benefit. This creates a form of sustainability debt that continues to grow as more data accumulates.

**Academic Research Community**
The academic research community must establish the knowledge foundation needed to address digital waste effectively, with several key research directions requiring immediate attention.

1. Researchers should develop comprehensive assessment methodologies for generative AI systems that incorporate long-term projections. New methodologies must account for how the environmental footprint of stored data might evolve over generations as technologies change, energy systems transform, and climate conditions shift. These assessments should model different scenarios for data accumulation and retention to provide policymakers with evidence regarding potential future trajectories. Quantitative research on multigenerational storage impacts is essential. Studies should model the cumulative environmental footprint of storing different types of AI-generated content over periods of 50-200 years, timeframes that better reflect the potential longevity of digital infrastructures. This research should account for projected changes in technologies, energy systems, and climate conditions to provide more accurate assessments of intergenerational burdens.

2. Researchers should establish normative frameworks for evaluating the justifiability of digital waste across generations, exploring questions like: What responsibilities do we have to future generations regarding the digital infrastructure we create? What constitutes fair distribution of benefits and burdens across generations? How should we balance present convenience against future environmental costs?

3. The psychological and social dimensions of generative AI usage require investigation, particularly regarding how users conceptualize the longevity of their digital content. Other questions that need to be explored are: What factors mitigate data generation volumes amongst users? How do users (re-)engage with generated data? Are users aware of the environmental impacts of data storage? Understanding these patterns (and others) will enable interventions that help users internalize the long-term impacts of their storage decisions.

4. Researchers should establish intergenerationally fair data lifecycle frameworks for generative AI content with clear protocols for determining the value of generated content over time and criteria for content retirement, archiving, or deletion. These frameworks should include mechanisms for periodic reassessment that allow future generations to participate in decisions about maintaining digital legacies rather than being bound by perpetual storage commitments made in the past.

**Developers**
Developers have a direct influence over the technical architecture and user experience of these technologies, positioning them to implement significant interventions that reduce digital waste at scale. These technical interventions represent an opportunity to embed sustainability values directly into system design.

1. File format optimization represents a foundational intervention. Developers should explore the creation of compression algorithms and file formats specifically designed for AI-generated content, recognizing unique characteristics that enable more efficient storage. Text-to-image systems might store the original prompt and seed value alongside a compressed output, enabling regeneration on demand rather than storing full-resolution images indefinitely.

2. Data lifecycle management should become a core feature rather than an afterthought. Developers should implement expiration protocols that suggest or automate content pruning after specified periods of non-use, potentially with graduated approaches that compress content after initial inactivity before suggesting deletion to users. These systems should provide clear information about storage implications while encouraging more sustainable practices through interface design.

3. Storage efficiency metrics would enable more transparent environmental assessment. Developers should create standardized measurements that quantify storage efficiency of different models, incorporating these metrics into evaluation frameworks alongside traditional performance measures. Making these metrics transparent would allow users to make informed choices based on environmental preferences.

4. System architecture should increasingly emphasize content reuse rather than regeneration. Developers should

build systems capable of efficiently modifying existing content rather than generating entirely new outputs for small changes. These approaches would reduce both computational resources required for generation and the storage burden of maintaining near-duplicate outputs, exemplifying the ethical principle of sufficiency.

5. Developers should implement "digital environmental impact" notifications within interfaces, providing users with real-time feedback about the environmental consequences of their generation and storage decisions. By making typically invisible impacts visible, such features could promote more conscious content management decisions among users.

**End-User and Organizations**

Sustainable management of generative AI content ultimately requires cultural and operational shifts among end-users and organizations. These shifts represent a move towards responsible digital stewardship, taking accountability for the environmental consequences of technological choices rather than externalizing these costs to society and future generations.

Organizations adopting generative AI should implement principles from established resource management approaches. This begins with comprehensive education to ensure employees understand both the capabilities of generative AI and the environmental implications of digital waste. Organizations should cultivate a culture that values data minimalism, perusing efficient storage practices rather than indiscriminate accumulation. This cultural development should precede large-scale deployment, establishing sustainable usage patterns from the outset rather than attempting to reform entrenched wasteful practices later.

Formal data governance policies provide essential structure for managing AI-generated content. These policies should establish guidelines for content retention based on utility, purpose, and regulatory requirements, articulating different tiers of storage with corresponding retention periods. They should assign clear responsibility for content management across the organization, ensuring that digital waste reduction becomes an ongoing operational priority rather than a one-time initiative. From an ethical perspective, these policies represent institutional commitments to environmental responsibility.

Regular data auditing practices should complement governance frameworks. Organizations should schedule periodic reviews of stored AI-generated content to identify unnecessary files and remove them from active storage. These audits might employ both automated tools and manual reviews that incorporate domain expertise regarding content value.

Organizations should develop frameworks for distinguishing high-value content from content with rapidly diminishing value. By clearly identifying which content merits long-term preservation, organizations can focus storage resources on maintaining truly valuable assets while allowing less critical content to expire naturally.

For individual end-users, education about the environmental impact of digital storage represents a critical intervention. Accessible educational materials that explain these connections could promote more conscious choices about content generation and retention.

## Digital Waste as an Ethical Imperative for Sustainable AI and Future Generations

As generative AI transforms content creation and consumption, we must broaden sustainability considerations beyond model training and inference to include the entire data lifecycle. Digital waste represents a significant environmental and ethical challenge with profound implications for future generations.

The intergenerational dimension requires particular emphasis. Unlike many environmental challenges that manifest immediately, the consequences of unmanaged digital waste accumulate gradually but inevitably. Each generation that fails to address this accumulation passes on an increasing burden to subsequent generations, who inherit not only the data but also the obligation to maintain supporting infrastructure. This creates environmental debt that compounds across generations, potentially becoming unmanageable if left unaddressed.

This paper contributes to AI sustainability research and AI development in several important ways. It expands responsible AI development to explicitly include environmental sustainability as a core consideration, with particular attention to intergenerational justice. By framing digital waste as an ethical issue with multigenerational implications, we establish that responsible AI development must account for the complete environmental footprint across extended timeframes.

The paper bridges disciplines by introducing concepts from digital resource management, lifecycle assessment, and intergenerational ethics to the discourse on ethical and sustainable AI. This interdisciplinary approach demonstrates that responsible AI development does not need to reinvent sustainability frameworks but can adapt established methodologies from adjacent fields.

We provide a structured foundation for addressing digital waste through specific recommendations that explicitly consider long-term impacts. These recommendations operationalize abstract ethical principles into concrete practices that can be implemented today to prevent accumulation of an unsustainable digital legacy.

This work expands the scope of AI ethics beyond its traditional focus on immediate harms to include considerations of intergenerational justice and global environmental impact. Just as decisions made by previous generations about industrial production created environmental challenges we face today, our digital practices will shape the world inherited by our descendants.

Addressing digital waste requires coordinated effort across the AI ecosystem. By introducing this concept as an ethical imperative rather than merely a technical challenge, we hope to inspire a more comprehensive approach to AI development that places sustainability at its core. As climate challenges intensify, the stakeholders involved in AI development have both the opportunity and responsibility to lead in sustainable digital practices that minimize environmental harm while maximizing human benefit across generations.